\documentclass[aps,pre,showpacs,twocolumn,superscriptaddress,floatfix]{revtex4-1}
\pdfoutput=1
\usepackage{graphicx,color,epsfig}
\usepackage[latin1]{inputenc}
\usepackage{booktabs} 
\begin{document}
\title{Complex networks vulnerability to module-based attacks}

\author{Bruno Requião da Cunha}
\affiliation{Instituto de Física, UFRGS, Brasil.}
\affiliation{Polícia Federal, Brasil.}

\author{Juan Carlos González-Avella}
\affiliation{Instituto de Física, UFRGS, Brasil.}

\author{Sebastián Gonçalves}
\affiliation{Instituto de Física, UFRGS, Brasil.}

\date{\today}

\begin{abstract}
In the multidisciplinary field of Network Science, optimization of
procedures for efficiently breaking complex networks is attracting
much attention from practical points of view.  In this contribution we
present a module-based method to efficiently break complex networks.
The procedure first identifies the communities in which the network
can be represented, then it deletes the nodes (edges) that connect
different modules by its order in the betweenness centrality
ranking list.  We illustrate the method by applying it to various well
known examples of social, infrastructure, and biological networks.  We
show that the proposed method always outperforms vertex (edge) attacks
which are based on the ranking of node (edge) degree or centrality,
with a huge gain in efficiency for some examples.  Remarkably, for the
US power grid, the present method breaks the original network of 4941
nodes to many fragments smaller than 197 nodes (4\% of the original
size) by removing mere 164 nodes ($\approx$ 3\%) identified by the
procedure. By comparison, any degree or centrality based procedure,
deleting the same amount of nodes, removes only 22\% of the original
network, {\it i.e.} more than 3800 nodes continue to be connected
after that.
\end{abstract}

\maketitle

\section{Introduction}\label{intro}
Network theory and its applications pervade many different scientific
fields, from physics to sociology, engineering, epidemiology,
mathematics and economy ---to cite a few.  In the context of network
science, three important concepts have received much attention
recently: interdependent graphs~\cite{Buldyrev2010,Brummitt2012},
communities (or modules)~\cite{fortunato2010community}, and robustness
of networks facing targeted attacks~\cite{valente2012network}. In the
present work we address and bring together these last two concepts.

The robustness of networks against failures, targeted attacks to
individuals components, and the impact on the performance of the
system has become an important issue for practical reasons in the last
few years.  In this sense, the robustness of a network is often
related to the structural functionality of the system as a whole, so
information can propagate over the network for example.  For instance,
the failure of routers in the Internet~\cite{Cohen2001}, the
vaccination of individuals to prevent the spread of a
disease~\cite{PastorSatorras2002}, and fighting organized crime or
terrorist groups~\cite{xu2008topology} can all be described by a
formal model in which a certain number of vertices (edges) in the
network are removed~\cite{iyer2013attack}.  Therefore, the robustness
of a complex network is directly related to the fraction of nodes
(edges) needed to be removed so that the network loses its
functionality.
%%%%%%%%%%%%%%%%%%%%%%%%%% F1 %%%%%%%%%%%%%%%%%%%%%%%%%%%%%%%%%%%%%
%\begin{figure}[t]
\begin{figure}
\begin{center}
\includegraphics[width=0.8\columnwidth]{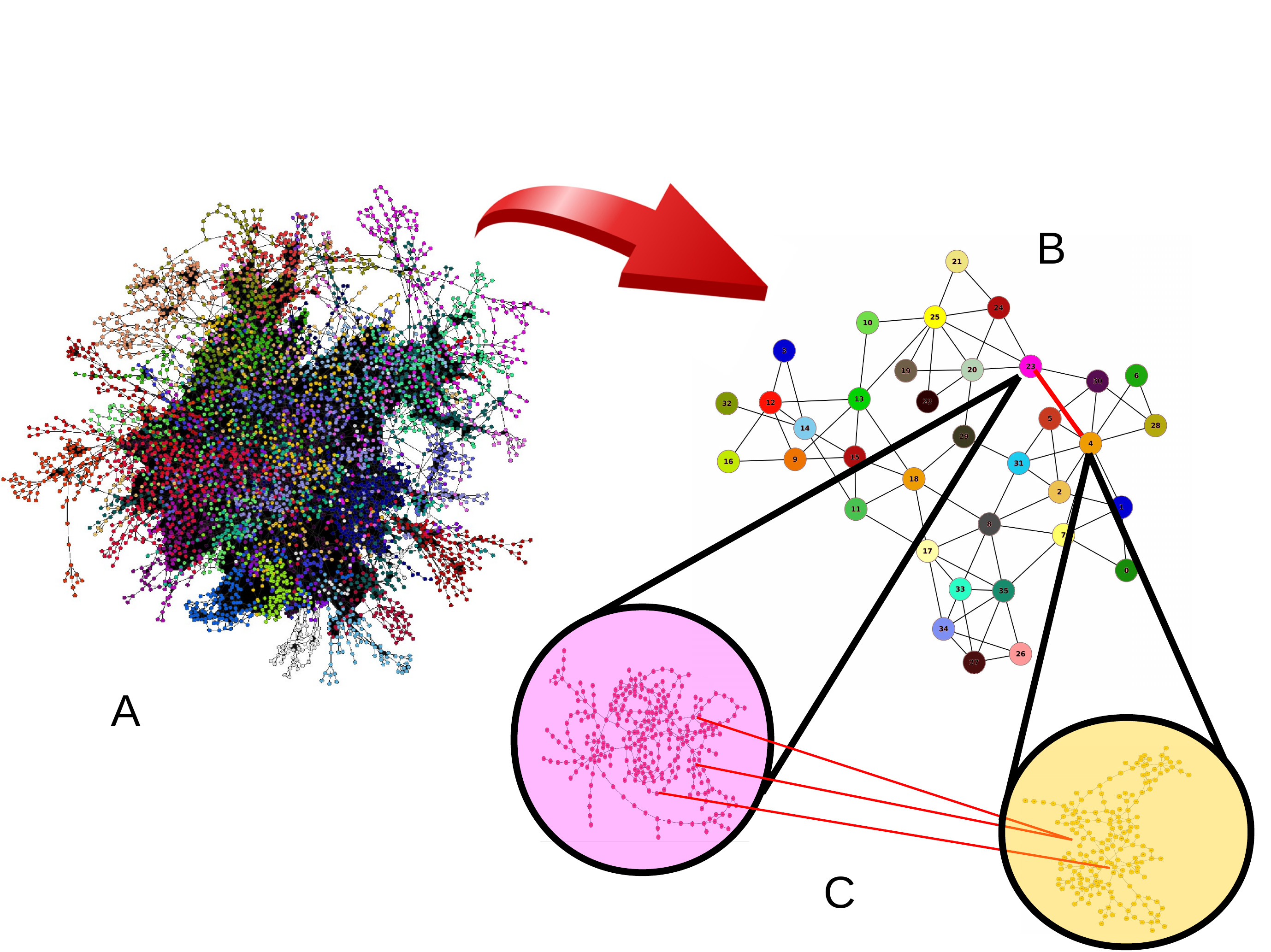}
\end{center}
\caption{Module based attack scheme: (A) represents the original
  network, (B) is one possible module representation of A, and (C)
  shows the internal structure of nodes and edges inside two selected
  modules and the edges connecting nodes between them; those edges
  (nodes) are the ones to be selected for deletion. The graph depicted
  in A corresponds the US power
  grid~\cite{konect2014opsahlpowergrid}.}
\label{mbafig}
\end{figure}
%%%%%%%%%%%%%%%%%%%%%%%%%%%%%%%%%%%%%%%%%%%%%%%%%%%%%%%%%%%%%%%%%%%%
Conversely, the less the number of nodes that a method identifies to
break down a network, the more efficient it is.  In this sense, many
centrality indexes have been proposed aimed to measure the structural
importance of nodes (edges)~\cite{iyer2013attack}.  For instance, the
concept of bridging nodes in the topology of complex networks has been
brought to discussion too~\cite{valente2010bridging}.  Hwang
\textit{et al.}~\cite{hwang2006bridging} define a bridging centrality
in order to characterize the location of nodes among high degree
nodes. The method succeeds in identifying functional modules but does
not show significantly better results than simple betweenness attack
when it comes to atomize different complex networks.  Nevertheless,
recent developments in efficient community extraction algorithms from
complex graphs~\cite{newman2006finding,blondel2008fast} show a
promising pathway in devising better attack strategies. In effect,
communities or modules are topological partitions of graphs with dense
internal connections but weakly connected among them. In this sense, in Fig.~\ref{mbafig} we depict the community structure for the Western United States Power Grid, illustrating this weak connection among clusters internally dense. Henceforth, a
natural question arises: How structurally important are those weak
interactions bridging distinct communities and how are they related to
the robustness problem? To answer this query is precisely the main
objective of this study. The work is organized as follows: section
~\ref{attack} discuss the generalities of attack in networks and the
concept of robustness, section ~\ref{mba} describes our method to
perform the attacks, while in section ~\ref{resu} we present the
results of the procedure to ten examples of real networks with
conclusions summarized in section ~\ref{conc}.

\section{Attacks}\label{attack}
In order to quantify the effect of the attacks on the networks~\cite{barthelemy2011spatial}, we
define $\mathcal{G}$ as an initial network of size $N$, and
$\mathcal{G}_\rho$ as the network that results after the removal of a
fraction $\rho$ of vertices (edges). Then we denote by
$\mathcal{L}_\rho$ the largest component of $\mathcal{G}_\rho$, whose
size we denote by $N_{\mathcal{L}}$. We define the order parameter
$\sigma(\rho)$=$\frac{N_{\mathcal{L}}}{N}$ which allows us to quantify
the response of a network as a function of the fraction of nodes
(edges) deleted.

An hypothetical way of getting the ordered list of targeted nodes to
be removed would be by brute force: try all the possible lists until
find the one that reduce the network to a desired size with the
minimum number of remotions. However this is useless because it means
checking $N!$ possible lists, which is computational prohibited for any
network bigger than $N \approx 12$.  On the other hand, the simplest
but no efficient strategy is random deletion of nodes, {\it i.e.}
make a random list of the nodes and remove them in that order. This
generally gives rise to a linear degradation of the network in which 
the fraction removed is mostly not much than the nodes removed.  A more
efficient and doable way of attack a graph consists in the deletion of
vertices (edges) in order of their importance in the structural
functioning of the network. In this sense, traditional attacks focus
on sorting nodes (edges) in decreasing order of some centrality index
---the so called Centrality-Based Attack (CBA)---, which perform much
better than random attacks.  Betweenness centrality, for instance,
takes a time, depending on the algorithm used, of only $O(N\times E)$,
where $E$ stands for the number of edges in the network.  This way, if
we choose some method as a null reference, the gain in efficiency can
be computed by the normalized ratio
\begin{equation}
\Gamma(\rho)=1-\sigma(\rho)/\sigma(\rho)_{nul}
\label{gamma}
\end{equation}
which increases as the attack method becomes more efficient than the
reference one.

Even though most attack methods focus on centrality ranking, real
networks tend to group into sparsely connected clusters and the
removal of few bridging structures should be able to detach large
chunks of densely connected nodes, leading to large values of
$\Gamma(\rho)$, as we shall see in the next section.

\section{Module-based attack}\label{mba}
The structural importance of a node (edge) depends both on local and
non-local measures. Hence, in the scope of the method proposed in this
paper, centrality and community detection are the topics that we
address to characterize and sort nodes (edges) in order to develop the
attack on networks.  As pointed out in the works by Iyer \textit{et
  al.}~\cite{iyer2013attack} and Holme \textit{et
  al.}~\cite{holme2002attack} nodes with high betweenness and high
degree are usually strongly correlated and both attacks have similar
efficiency.  Besides, previous work shows that for real networks the
betweenness-based method is in general the most
efficient~\cite{iyer2013attack}. Thence, from now on we take
betweenness centrality attack as our reference or null method.

Likewise, vertices connecting different communities generally have
high betweenness centrality since many shortest paths pass through
them. However, as fewer connections are expected among communities
these nodes are not the ones with higher degree.  Therefore, in order
to detach communities in a very efficient way, we propose a
Module-Based Attack (MBA) consisting of sorting all nodes (edges) by
betweenness centrality, then choosing only those nodes (edges) that
link different communities. One should note that in vertex attack, as
we aim to detach previously detected communities, once a node from a
bridging edge is deleted, there is no need to detach its counterpart
unless it also participates in other inter-communities
connections. Besides, at each step of the procedure, the attack will
focus on the remaining largest connected component of the network, in
order to speed up the fragmentation.  This process loosely resembles
the original idea of weak ties proposed by
Granovetter~\cite{granovetter1973strength} for social networks and
later developed in the framework of topological communities by De Meo,
Ferrara \textit{et al.}~\cite{de2014facebook}.

%%%%%%%%%%%%%%%%%%%%%%%%%% F2 %%%%%%%%%%%%%%%%%%%%%%%%%%%%%%%%%%%%%
\begin{figure}[!]
\begin{center}
\includegraphics[width=0.9\columnwidth]{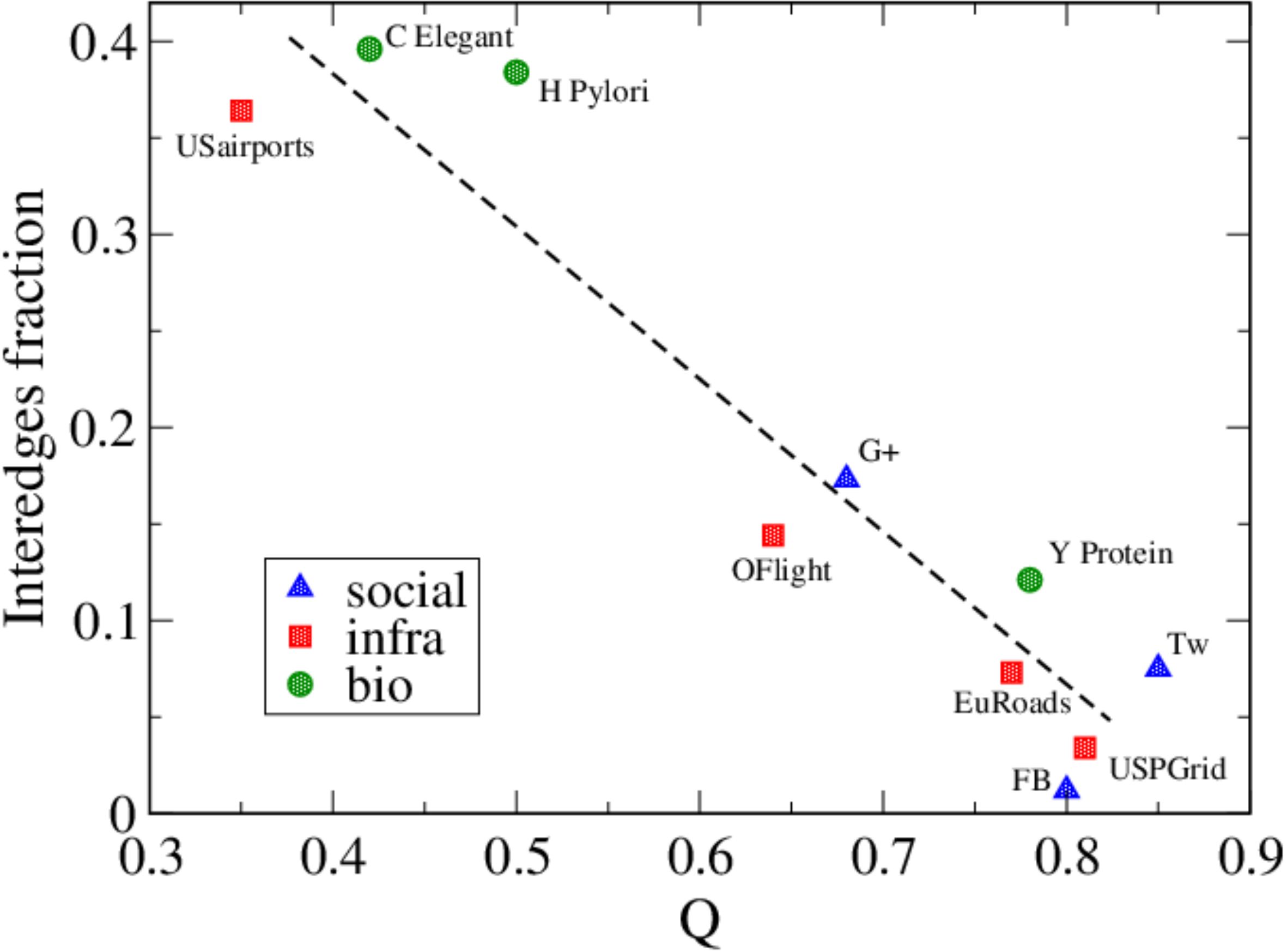}
\end{center}
\caption{Fraction of edges bridging communities vs modularity $Q$ for
  the ten real networks studied in this work. The dash line is the
  resulting linear fitting: $y = 0.7 - 0.8x$, with a correlation
  coefficient $R=-0.95$.}
\label{edgeq}
\end{figure}
%%%%%%%%%%%%%%%%%%%%%%%%%%%%%%%%%%%%%%%%%%%%%%%%%%%%%%%%%%%%%%%%%%%%

\begin{table*}[tp]
\setlength{\arrayrulewidth}{.05em}
  \setlength{\tabcolsep}{2.5pt}
       \begin{tabular}{lrrrrrrrrrrrrr}
    \hline

    \toprule
      & $\langle Q\rangle$ & $\langle k \rangle$&$\langle k^{2}\rangle$ & $N$ & 
   $E$ &$\bar{N}_{mod}$ & $N^{max}_{mod}$ & $E_{inter}$ & $\eta_{E}$ & $\eta_{N}$\\ \hline

    Facebook & 0.80 & 2.06  & 528 & 2888 & 2981& 8 & 0.260 & 0.012 & 0.49 & 0.12\\ [0.1cm]%\hline
    Twitter & 0.85 &  2.81 & 108 & 23370 & 32831& 136 & 0.056 & 0.075 & 0.84 & 0.36\\ [0.1cm]%\hline  
    Google Plus & 0.68 & 3.32 & 1250 & 23628 & 39194& 33 & 0.144 & 0.173 & 0.65 & 0.47\\ [0.1cm]\hline
   
    US Power Grid & 0.81 & 2.67 & 10 & 4941 & 6594 & 40 & 0.040 & 0.034 & 0.86 & 0.77\\[0.1cm] %\hline
    Euro Road & 0.77 & 2.41 & 7 & 1174 & 1417 & 47 & 0.061 & 0.073 & 0.79 & 0.71\\ [0.1cm]%\hline
    Open Flights & 0.64 & 10.67 & 594 & 2939 & 15677& 38 & 0.144 & 0.144 & 0.54 & 0.32\\ [0.1cm]%\hline    
    US Airports & 0.35 & 21.87  & 2454 & 1574 & 17215& 11 & 0.311 & 0.364 & 0.06 & 0.22\\ [0.1cm]\hline
    
    Yeast Protein & 0.78 & 2.39 & 15 & 1846 & 2203 & 179 & 0.068 & 0.121 & 0.59 & 0.55\\ [0.1cm]%\hline
    H Pylori & 0.50 & 3.88 & 45 & 724 & 1403& 25 & 0.097 & 0.384 & 0.45 & 0.30\\ [0.1cm]%\hline
    C Elegans & 0.42 & 8.94 & 358 & 453 & 2025& 9 & 0.232 & 0.396 & 0.37 & 0.20\\ %\hline
    \bottomrule \hline
    \end{tabular}
    \caption{Topological data for several networks consisting of
      average modularity, mean degree, second momentum, size of the
      original network, total number of edges, mean number of modules,
      relative size of the largest community, fraction of edges
      linking distinct communities, and relative overall efficiency of
      MBA method as compared to betweennss-based method for edge and
      node attacks.}
\label{table}
\end{table*}

Nonetheless, this procedure is slower than traditional methods since
fast community extraction takes, depending on the algorithm used,
about $O(N^2)$, resulting in a lower limit of computation of
$O(N^2+N\times E)> O(N\times E)$, where $N$ is the total number of
nodes and $E$ stands for the total number of edges.

As a preemptive measure of our proposed method we show in
Fig.~\ref{edgeq} the relation between the fraction of nodes in the
interface of communities, {\it i.e.} the fraction of nodes that make
the connection between the different modules extracted from the
networks, and the value of the modularity for each one of the real
networks that we present in the next section.  As expected, we observe
an approximate linear (negative) correlation between the fraction of
edges bridging communities and the modularity $Q$, which is precisely
the desired feature that makes the method potentially well posed. 
As we can see, the infrastructure networks are the ones which better adjust to the linear
behavior while the social networks are the worst cases.

\section{Results}\label{resu}
We now apply the method to real networks with different topological
structures. We investigate the behavior of such systems when
topological characteristics are measured only once before the
attacking procedure ---the so-called simultaneous attack.  Besides,
the graphs were taken as undirected.  For the networks studied here,
the different methods for communities extraction, {\it i.e.}
multilevel~\cite{blondel2008fast}, fast
greedy~\cite{clauset2004finding}, walktrap~\cite{pons2005computing},
infomap~\cite{rosvall2009map}, and leading
eigenvector~\cite{newman2006finding}, have all a community membership
coincidence higher than $90\%$.  Thence, in these simulations we have
used the method proposed by Blondel \textit{et
  al.}~\cite{blondel2008fast} because it is the quickest in computing
time.

%%%%%%%%%%%%%%%%%%%%%%%%%% F3 %%%%%%%%%%%%%%%%%%%%%%%%%%%%%%%%%%%%%
\begin{figure*}[tp]
\begin{center}
\includegraphics[width=1\linewidth,angle=0]{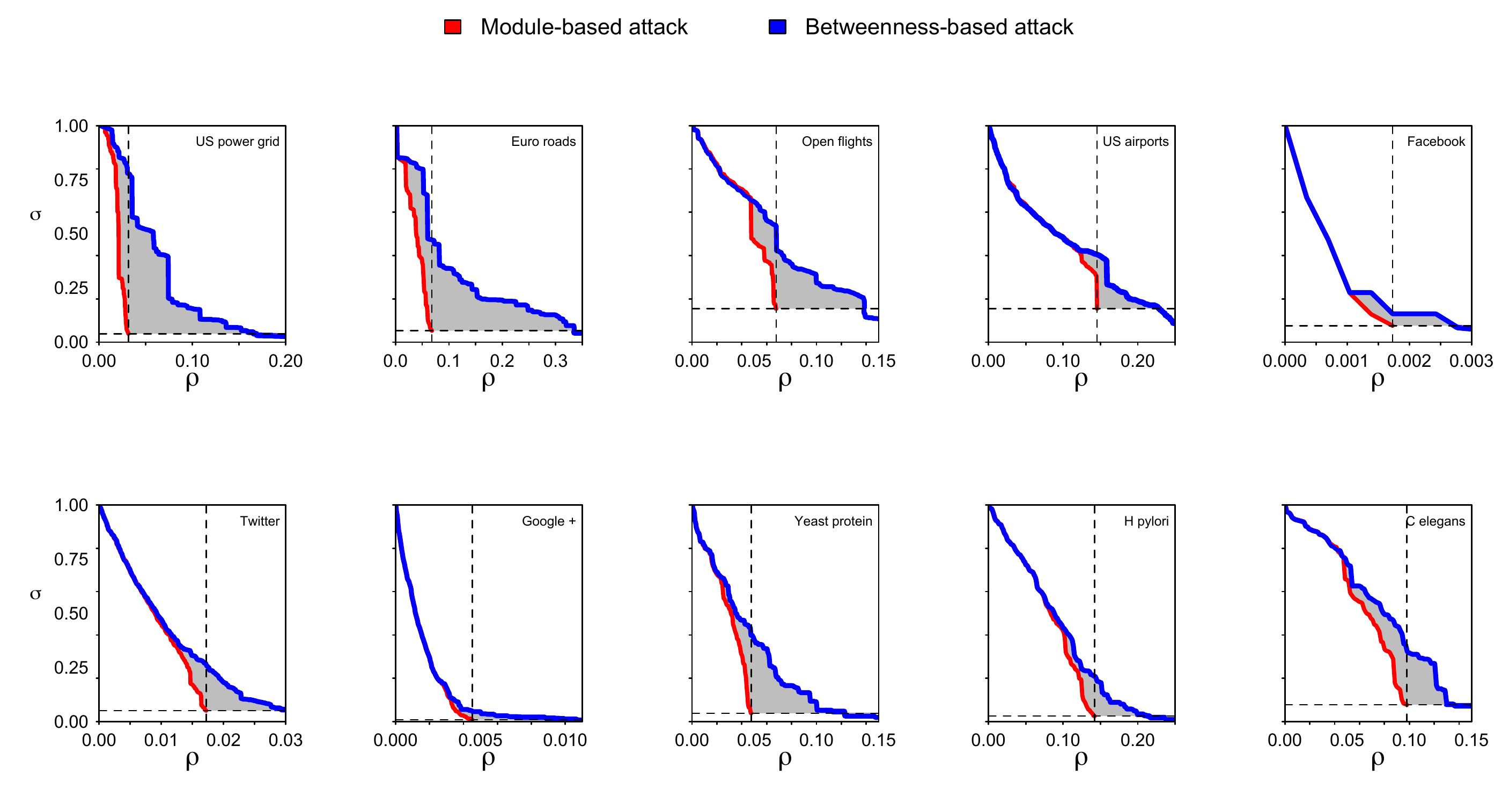}
\end{center}
\caption{Vertex MBA applied to real networks. Figure portrays the size
  of the biggest connected component relative to the original
  network's size, $\sigma$, in function of the fraction of removed
  nodes, $\rho$. Network data are explained in Table~\ref{table}.}
\label{vertex}
\end{figure*}
%%%%%%%%%%%%%%%%%%%%%%%%%%%%%%%%%%%%%%%%%%%%%%%%%%%%%%%%%%%%%%%%%%%%

%%%%%%%%%%%%%%%%%%%%%%%%%% F4 %%%%%%%%%%%%%%%%%%%%%%%%%%%%%%%%%%%%%
\begin{figure*}[tp]
\centering
\begin{center}
\includegraphics[width=1\linewidth,angle=0]{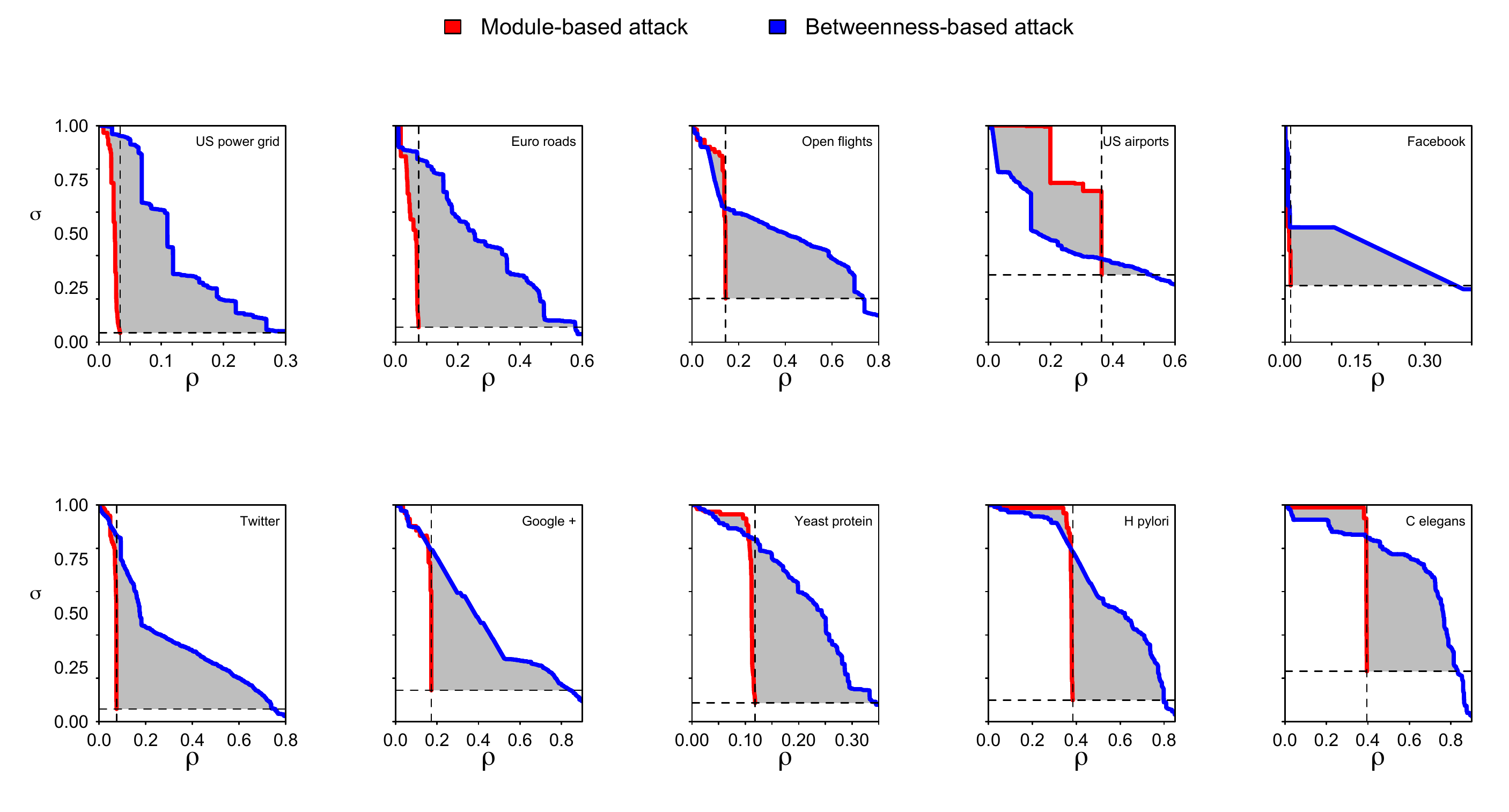}
\end{center}
\caption{Edge MBA applied to real networks. Figure portrays the size
  of the biggest connected component relative to the original
  network's size, $\sigma$, in function of the fraction of removed
  links, $\rho$. Network data are explained in Table~\ref{table}.\label{edges}}

\end{figure*}
%%%%%%%%%%%%%%%%%%%%%%%%%%%%%%%%%%%%%%%%%%%%%%%%%%%%%%%%%%%%%%%%%%%%

We have chosen three distinct groups of networks: infrastructure (US
Power Grid, Euro Road, Open Flights and US
Airports)~\cite{konect2014opsahlpowergrid, konectduncan98,
  konect2014subeljeuroroad, konecteroads,
  konect2014opsahlopenflights, konectopsahl2010b,
  konect2014opsahlusairport, konectopsahl11}, biological (Yeast
Protein, C Elegans and H Pylori)~\cite{jeong2001lethality,
  rain2001protein, konect2014arenasmeta, konectduch05} and social
(Facebook, Google+ and Twitter)~\cite{konect2014egofacebook,
  konect2014egogplus, konect2014egotwitter, konectMcAuley2012}.
In the Euro Road network, nodes represent European cities and edges
represent roads. Power Grid stands for the electrical power grid of
the Western States of the United States of America.  An edge
represents a power supply line and a node is either a generator, a
transformer, or a substation.  The Yeast Protein interaction network
is the same as in ~\cite{jeong2001lethality}. In the metabolic network
of the roundworm \emph{Caenorhabditis elegans} nodes are metabolites
(e.g., proteins) and edges are interactions between them.  The
\emph{Helicobacter pylori} is the same protein-protein interaction map
as in ~\cite{rain2001protein}.  In the Facebook user-user friendship
network (NIPS) nodes represent users and edges represent friendship.
Similarly, in the Google+ network, an edge means that one user (node)
has the other user (node) in her/his circles, while in the Twitter
network an edge indicates that both users (nodes) follow each other.

Simulations show that vertex MBA always outperforms the traditional
betweenness attack. Initially, both methods are similar but, as
bridging nodes are deleted, whole communities start to detach from the
core of the graph, resulting in large atomization of the network and
hence in an abrupt increase of $\Gamma$. On the other hand, in edge
MBA the situation changes. As we erase solely edges bridging modules,
initial attacks result in a plateau in $\sigma$ until we effectively
detach whole modules. After this critical point is reached, $\sigma$
decreases abruptly, relatively large communities are detached
extremely fast, and the whole network falls apart. 
Attacks usually stop before $\sigma \rightarrow 0$, depending on the
particular modular structure of each network, at a point
$P_{c}=(\sigma_{c}, \rho_{c})$. This happens precisely when all
original communities are detached with no targeted node (edge) left in
the remaining clusters, so the network stops functioning as a whole
---for instance, information would be stacked within the communities and
these structures would not be able to communicate each other.  From
this point on, one may continue to strike the network (or what is
left of it, which is the remaining largest connected component, generally
speaking, of the same size of the biggest original community) by some
classical attack based on centrality measures such as degree or
betweenness.  Looking at the figures of the attacks, we can say that
vertex MBA shows a second-order phase transitions type behavior
(Fig.~\ref{vertex}), while edge MBA exhibits a typical first-order
phase transition behavior (Fig.~\ref{edges}). In either case, with
$\sigma(\rho)$ as the order parameter, the critical points shown in
Figs.~\ref{vertex} and ~\ref{edges} mark what we call modular
percolation, \textit{i.e.} the point at which the network is modularly
disconnected.

%%%%%%%%%%%%%%%%%%%%%%%%%% F6 %%%%%%%%%%%%%%%%%%%%%%%%%%%%%%%%%%%%%
\begin{figure}[bp]
\begin{center}
\includegraphics[width=1\linewidth,angle=0]{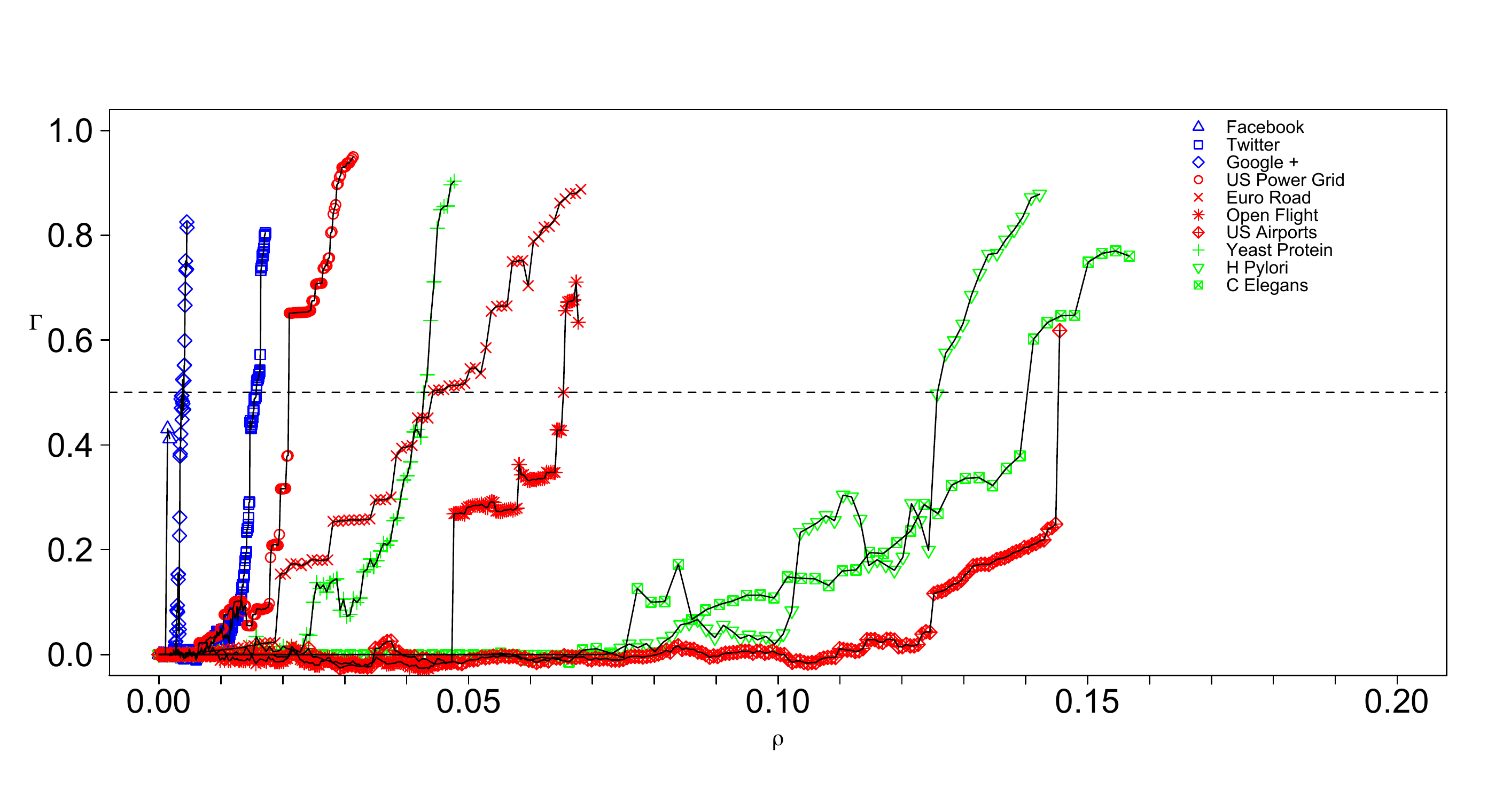}
\end{center}
\caption{Gain in efficiency of vertex MBA, compared to vertex CBA, as
  a function of the fraction of removed nodes. Network data are
  explained in Table~\ref{table}.}
\label{gain}
\end{figure}
%%%%%%%%%%%%%%%%%%%%%%%%%%%%%%%%%%%%%%%%%%%%%%%%%%%%%%%%%%%%%%%%%%%%

In general, attacking nodes is more efficient than edges since the
removal of a vertex always results in the deletion of all edges
attached to it. However, depending on the real system studied, vertex
or edge attack may not make sense. For instance, in the case of Euro
Road one may envisage blocking the traffic between two cities, while
removing a node would mean to erase an entire village. On the other
hand, in biological systems for example, node deletion makes sense,
since individual metabolites are susceptible to be removed
from the network.

With these results we now plot the efficiency $\Gamma$ of the MBA as compared to CBA as a function of $\rho$ for each network in Fig.~\ref{gain}. It is easy to
see that most networks reach a gain in efficiency of more than 50\% for less than 7\% of nodes removed -- the more oustanding case being 
the US Power Grid with more than 95\% of gain with approximately 3\% of nodes removed. Even in the worst cases (Yeast Protein, H pylori and US Airports) we get more than 60\% of gain for less than 16\% of vertices deleted.

\begin{figure*}[tp]
\begin{center}
\includegraphics[width=0.9\linewidth]{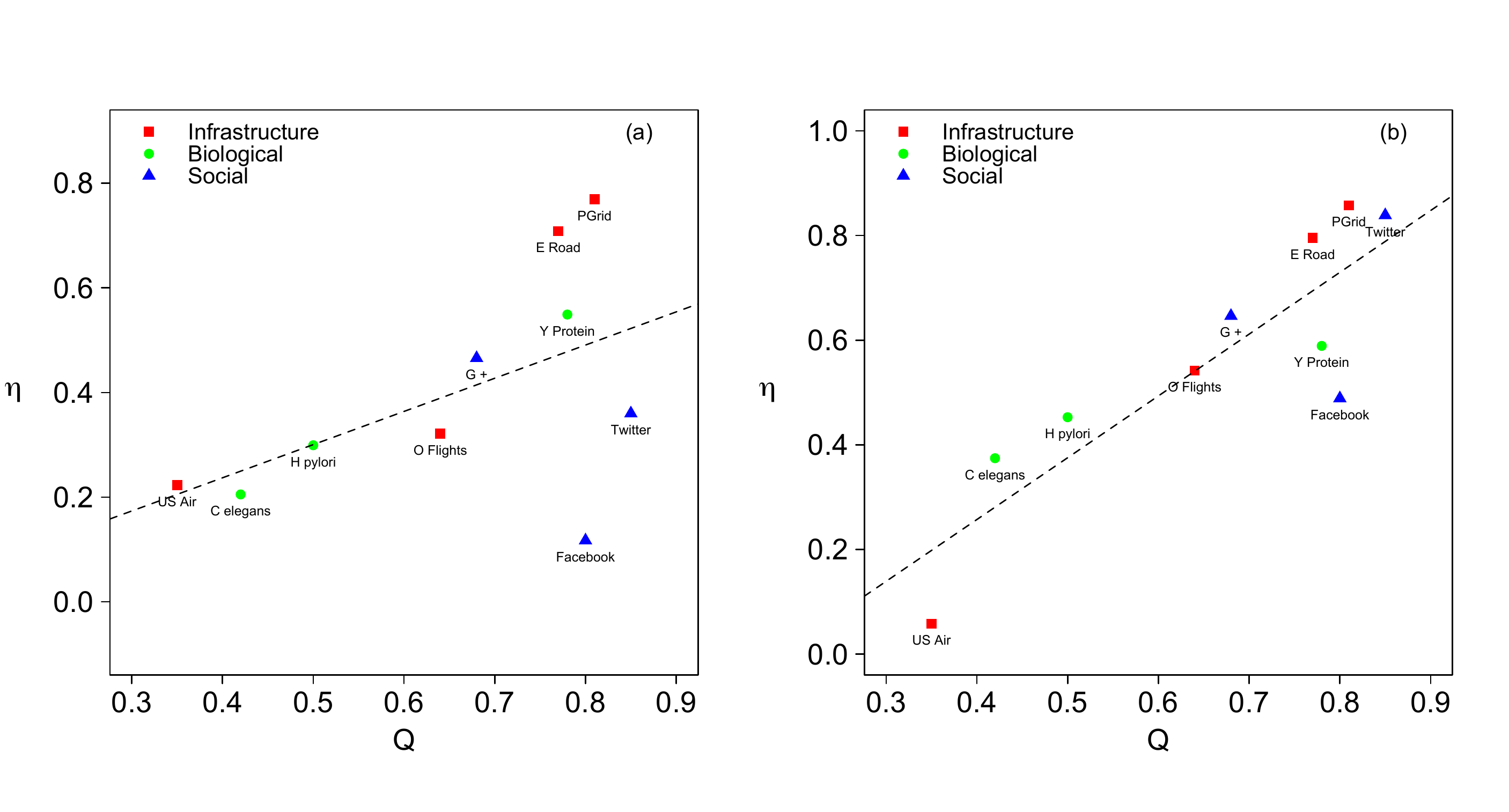}
\end{center}
\caption{Overall efficiency $\eta$ of MBA, compared to CBA, as a
  function of modularity for (a), nodes and (b), edges removal.}
\label{eta}
\end{figure*}

The overall efficiency of MBA as compared to CBA may be measured by
how fast MBA reaches the modular critical point relative to
CBA. In other words, we can define the relative overall efficiency as:
\begin{equation}
 \eta=\Gamma_{c}\times\varrho_{c}
\end{equation}
where $\Gamma_{c}$ is calculated at the critical point ($\rho_c$) and
$\varrho_{c}$ is defined as $1-\rho/\rho_{null}$ where $\rho$ stands
for the fraction of nodes removed at the critical $\sigma$ in the MBA
approach and $\rho_{null}$ is evaluated at the same y-axis point but
in the CBA curve.  As before, the reference or null method of attack
is betweenness-based. The results for the present method of attack on
the ten networks, in terms of the previous defined $\eta$, shown in
Fig.~\ref{eta}, tells us that its efficiency strongly depends on the
modularity.  This is an expected phenomenon since networks with high
modularity tend to have a density of edges connecting communities much
smaller than the density of internal edges. 

%%%%%%%%%%%%%%%%%%%%%%%%%% F7 %%%%%%%%%%%%%%%%%%%%%%%%%%%%%%%%%%%%%
\begin{figure}[h]
\begin{center}
\includegraphics[width=0.6\linewidth]{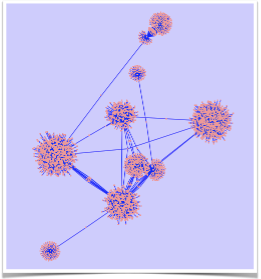}
\end{center}
\caption{Facebook sub-graph depicting its trivial modular structure.}
\label{facebook}
\end{figure}
%%%%%%%%%%%%%%%%%%%%%%%%%%%%%%%%%%%%%%%%%%%%%%%%%%%%%%%%%%%%%%%%%%%

However, in the node
removal approach, we have a slightly different picture since the
attacks also break down the internal structures of the modules,
departing from a linear fit. There is also another important effect
that should be noticed: some networks have high modularity, but the
inner structure of the modules is very weak. In these cases, the MBA
approach does not introduce major gains in efficiency, even though it
is still more efficient than traditional CBA. The fact is that in
these systems the inner community structure is weak so a more simple
attack can be as efficient as the one presented here.  On the other
side, we may have networks with smaller modularity, but for which the
impact of the MBA approach is higher than the impact on networks with
higher modularity. That is precisely the special case of the Facebook
network extract studied here. As we can see from Fig.~\ref{facebook}
the community structure of this network is trivial, with most of the
bridging nodes corresponding to the ones with higher degree.  Besides,
the internal structures of modules are extremely weak with all nodes
connected only to a few vertices or even to only one central node.

%%%%%%%%%%%%%%%%%%%%%%%%%% F5 %%%%%%%%%%%%%%%%%%%%%%%%%%%%%%%%%%%%%
%%%%%%%%%%%%%%%%%%%%%%%%%%%%%%%%%%%%%%%%%%%%%%%%%%%%%%%%%%%%%%%%%%%%

%%%%%%%%%%%%%%%%%%%%%%%%%% F8 %%%%%%%%%%%%%%%%%%%%%%%%%%%%%%%%%%%%%
\begin{figure*}[t]
\begin{center}
\includegraphics[width=1\linewidth,angle=0]{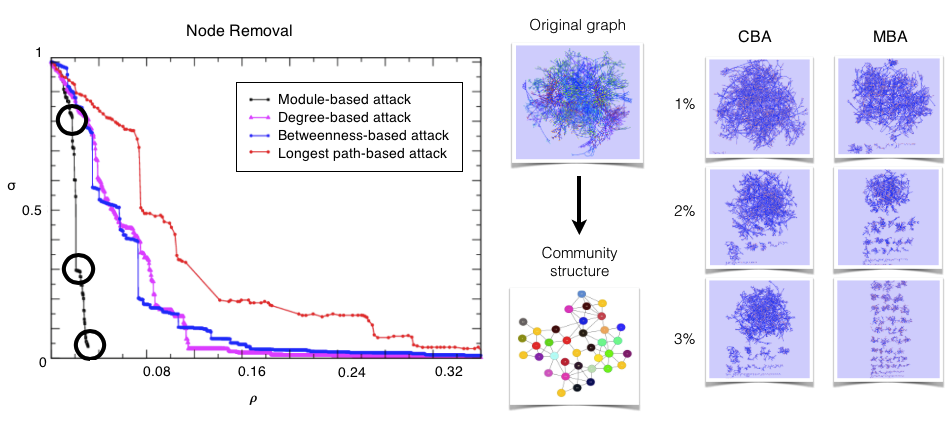}
\end{center}
\caption{Pictorial comparison between the effect of betweenness-based
  attack, degree-based attack, longest path
  attack~\cite{pu2015vulnerability} and module-based attack in the
  case of US Power Grid system for 1\%, 2\% and 3\% of nodes removed.}
\label{uspg}
\end{figure*}
%%%%%%%%%%%%%%%%%%%%%%%%%%%%%%%%%%%%%%%%%%%%%%%%%%%%%%%%%%%%%%%%%%%

Before arriving at the conclusion, we illustrate on the attack procedure
with a case where its performance is remarkably better than previous
and well accepted attacking prescriptions. That example is the
Power grid of Western USA. In this sense, Fig.~\ref{uspg} summarizes the result of 
our method of attack as compared to betweenness
centrality attack, degree centrality attack, and longest pathway
attack~\cite{pu2015vulnerability}, along with the snapshots of the 
network when 1\%, 2\% and 3\% of nodes are removed by betweenness
centrality ranking (CBA) and by the module-based method (MBA).
Remarkably, the present method breaks the original network of 4941
nodes to many fragments smaller than 197 nodes (4\% of the original
size) by removing mere 164 nodes ($\approx$ 3\%) identified by the
procedure. By comparison, in any degree or centrality based procedure,
deleting the same amount of
nodes, removes only 22\% of the original network, {\it i.e.} more than
3800 nodes continue to be connected after that. %In this sense,
% in Fig.~\ref{uspg} we represent the comparison among betweenness-based,
% degree-based, longest path-based and module-attacks.  
Such extreme atomization of the network is represented graphically 
by the set of figures on the far right of Fig.~\ref{uspg}. Besides, it is promptly seen that the community structure of this network is far from trivial.

\section{Conclusions}\label{conc}
We have presented a module-based attack method which consists of
erasing only those structures that bridge distinct communities ordered
by betweenness centrality.  Computational simulations on many real
networks show that the MBA method is more efficient in atomizing
networks than traditional procedures based only on centrality
criteria.  Henceforth, one may say that, in general, the most
connected or most central nodes are not necessarily the most important
for the network survival. Conceptually speaking, nodes (edges) linking
distinct communities are structurally more important and crucial for
the modular cohesion of the network than nodes (edges) with high
degree or centrality indexes.  Obviously, networks with high
modularity tend to be in general more fragile against module-based attacks, while
networks with intra-module weakness can show smaller differences
between different attacking methods.

It should be noted that, regarding the aim of the present work, the
communities that emerge from the networks by using the module
identification algorithms, have in principle no relation with real
communities.  However, the organization of a network in coarse grain
agglomeration may disclose important information about the structural
functionality of complex networks.

In discussions of community detection algorithm, the resolution limit
is a topic of debate, however in connection with the attack method
proposed here, it is not highly relevant. In fact, what is desirable is a
compromise solution in terms of the average module size and the
network size.  Large modules means a network decomposed in few of
them, which is good because many nodes are disconnected once a module
is detached from the others. The drawback is that the last module
could be still a large part of the original network. On the other
hand, a decomposition in many small communities has the advantage of
ending with a highly fragmented network, but at the expense of being
slower than the other scenario. Therefore the optimum is a
situation somehow in the middle, and that is the reason the resolution
limit of communities is not of high concern here.

As a final remark, we emphasize that eventual applications of
module-based attacks to classes of real systems such as terror, crime
or disease related networks might lead to groundbreaking procedures to
fight these unwanted threats.  We acknowledge CNPq and the Brazilian
Federal Police for financial support.

\bibliography{bib_paper1}
\end{document}